# Bridging the Landau theory of crystallization and the cluster approach to quasicrystals

O. V. Konevtsova[1], S. B. Rochal[1] and V.L. Lorman[2]

[1]*Faculty of Physics, Southern Federal University, 5 Zorge str., 344090 Rostov-on-Don, Russia*

[2]*Laboratoire Charles Coulomb, CNRS – Université Montpellier 2, pl. E. Bataillon, 34095, Montpellier, France*

We propose the theory which unifies the description of quasicrystal assembly thermodynamics and quasicrystal structure formation by combining the Landau theory of crystallization and the cluster approach to quasicrystals. The theory is illustrated on the example of pentagonal Penrose quasilattice. The coordinates of the quasilattice nodes are calculated by minimizing the Landau free energy with the constraint imposed by internal organization of atomic clusters, without explicit use of high-dimensional crystallography. The procedure proposed establishes direct relations between the cluster organization, quasicrystalline structure and thermodynamic properties of the quasicrystalline state. The correspondence is shown between the basic features of the proposed algorithm for quasilattice construction, on the one hand, and the conventional projection method, on the other hand. It provides a new physical justification for high-dimensional crystallography methods application.

PACS numbers: 61.50.Ah, 61.44.Br, 64.70.dg

## I. INTRODUCTION

The last decade has seen a considerable regain of interest to materials characterized by high positional order but noncrystallographic rotational symmetry and quasiperiodic translational properties [1]. In addition to classical quasicrystals (QC's), metallic alloys with surprising properties discovered already in 1984 [2], recent progress in soft and biological matter science and in metamaterials has opened access to a whole series of new quasiperiodic systems. Among them new dodecagonal QC's and their approximants in micellar systems [3], star polymers [4] and linear terpolymers [5], artificial QC's of laser-trapped particles [6], and quasicrystalline arrangements of proteins in 2D viral capsid shells [7]. Exceptional structures and physical properties of these new systems have induced a new wave of theoretical models [1] proposed to explain the peculiarities of quasicrystalline state of matter. Emergence of new QC's and their diversity determines new requirements to theoretical works: the relations between QC-forming structural units, global quasicrystalline structures, their thermodynamic stability and physical properties should be established in the same model frame [8]. However, historically, the understanding of the quasicrystalline state is developing in several different directions with not too many bridges between them.

Purely structural, crystallographic properties and peculiarities of diffraction patterns exhibited by QC's were understood using the concept of N-dimensional crystallography proposed initially by Wolf for modulated phase in 1974 [9] and appeared to be suitable for QC's too. Within this approach the structures are described by N-dimensional space groups spanning corresponding N-dimensional spaces, which are periodically decorated by atomic surfaces (AS's) [10]. Intersection of these surfaces with a cut plane generates atomic positions in the real structure.

The relation between quasicrystalline order and local structure of atomic clusters were studied in the frame of less rigorous from the crystallographic point of view, but equivalent and simpler description of real QC structures, proposed by Burkov [11]. According to Burkov, the QC structure is formed from equivalent building blocks (clusters), situated in the nodes of a quasilattice. The clusters in Burkov's model are overlapped, but the number of different overlaps is limited. In his original work, Burkov used the classical rhombic Penrose tiling [12] as a model quasilattice (QL). Burkov's model, applied to $Al_{65}Cu_{20}Co_{15}$ structure, was improved by Yamamato [13], who decorated with the same clusters the pentagonal Penrose tiling P1 (in the following consideration this tiling is called the pentagonal Penrose quasilattice (PPQ)). The cluster approach became a basis for many structural models of QC's [14] relating the global order with the cluster properties, and has even been formalized in a software package for structure analysis of QC's [15]. The cluster approach has also motivated more rigorous (but more abstract) models of plane coverings with overlapping clusters [16]. It was shown that overlap rules can be chosen so as to obtain a unique atomic arrangement isomorphic to a Penrose tiling. The thickness of covering, related to the degree of overlap, links then the local cluster organization to the global QC structure. However, these works do not deal explicitly with the QC stability justification or with the physical properties of QC's.

The study of physical properties of QC's has constituted an important, but practically independent from purely structural studies, field of research. A number of results in this field were obtained in the frame of the Landau theory of crystallization, based on the invariance properties of free energy functionals. A series of models explaining the stability of quasicrystalline state were proposed [17]. The explicit form for harmonic phonon-phason elastic energy in QC's with various symmetry was obtained [18]. It formed the basis for models for diffuse scattering in QC's [19], for the analysis of motion equations and for the study of phonon and phason modes coupling [20]. A detailed classification of defects in QC was also proposed [21]. The list of all significant results obtained in this frame would take several pages.

The description of new complex QC's and a new light shed on classical metallic quasicrystalline alloys need a unified vision of the quasiperiodic physical system. This implies the merging of (at least) three main parts of physics of QC's mentioned above. For that aim we propose in the present work to apply the Landau theory of crystallization to elucidate the peculiarities of QC structure formation. To construct a bridge between thermodynamic stability and physical properties, on the one hand, and both local and global structural properties of the QC, on the other hand, we combine the Landau theory with the cluster approach to the QC structure. In contrast to a conventional theory of crystallization, we consider the structure formation as a self-assembly of atomic clusters into a QL. The theory developed is illustrated on the example of PPQ which has several evident advantages, like simplicity and certain prevalence in experimental systems. It can be found not only in metallic alloys but in soft QC's and in protein nanoquasicrystals [7] also. We demonstrate that, in the frame proposed, the coordinates of the PPQ nodes are obtained by minimization of the Landau free energy with the constraint imposed by internal organization of atomic clusters. Corresponding effective free energy is shown to be dependent on the order parameter spanning single irreducible representation of the symmetry group of the isotropic state. The procedure does not use explicitly the methods of high-dimensional crystallography. Finally, by comparative analysis of QC structures we establish the correspondence between the basic features of the proposed algorithm and of the conventional N-d projection method, thus providing a new physical justification for high-dimensional crystallography methods application.

The Landau free energy $F$ is usually expanded in powers of amplitudes $\rho_k$ of the critical density waves system [22]. In contrast to the standard crystallization theory we take into account explicitly the dependence of $\rho_k$ values on the coordinates of cluster centers of mass. This fact allows us to apply the Landau theory for the calculation of the QL nodes coordinates. The application of the theory to the case of decagonal QC's is considered in detail in the following section. In Sec. III the minimization of the Landau free energy is performed with the constraints imposed by the simplest limitations on the possible distances between neighboring clusters, related, in turn to the internal structure of the clusters. The relation between the theory proposed and the N-d crystallography methods is illustrated by a direct comparison of the resulting algorithm with the conventional projection method.

## II APPLICATION OF THE LANDAU CRYSTALLIZATION THEORY TO THE STRUCTURE DETERMINATION

Let us recall some basic features of the conventional Landau crystallization theory [22] applied to planar decagonal quasicrystals [18]. Near a crystallization point the distribution density of structural units can be written as

$$\rho(\mathbf{R}) = \rho_0 + \delta\rho(\mathbf{R}), \qquad (1)$$

where $\rho_0$ is the cluster density before the crystallization and $\delta\rho(\mathbf{R})$ corresponds to a critical density deviation induced by a quasicrystalline order formation. According to the Landau theory $\delta\rho(\mathbf{R})$ is an irreducible function spanning single irreducible representation of isotropic state symmetry group. The corresponding plane-wave expansion takes the form:

$$\delta\rho(\mathbf{R}) = \sum_{k=0}^{9} \rho_k \exp(i\mathbf{b}_k \mathbf{R}), \qquad (2)$$

where $\mathbf{R}$ is a radius vector, $\mathbf{b}_k = b^0 \langle \cos(k\pi/5), \sin(k\pi/5) \rangle$. Since amplitudes $\rho_k = |\rho_k| \exp(i\phi_k^0)$ are taken in a complex form, and the density deviation $\delta\rho(\mathbf{R})$ is real, $\rho_\mathbf{k} = \rho_j^*$ for $j=(k+5)$ mod 10 and the number of independent phases $\phi_k^0$ is equal to 5. The simplest expansion of the Landau free energy is then expressed as the following invariant function of $\rho_\mathbf{k}$ amplitudes:

$$F = F_1 I_1 + F_2 I_2 + F_3 I_1^2 + F_4 I_3 + F_5 I_2 I_1 + F_6 I_1^3, \qquad (3)$$

where $F_i = F_i(P,T)$ are phenomenological coefficients dependent on temperature $T$ and pressure $P$; $I_1 = \rho_0\rho_5 + \rho_1\rho_6 + \rho_2\rho_7 + \rho_3\rho_8 + \rho_4\rho_9$, $I_2 = \rho_0\rho_1\rho_5\rho_6 + \rho_1\rho_2\rho_6\rho_7 + \rho_2\rho_3\rho_7\rho_8 + \rho_3\rho_4\rho_8\rho_9 + \rho_4\rho_5\rho_9\rho_0$, $I_3 = \rho_0\rho_2\rho_4\rho_6\rho_8 + \rho_1\rho_3\rho_5\rho_7\rho_9$ are basic invariant functions of density amplitudes $\rho_k$.

The minimal form of the Landau free energy is defined by its dependence on the density wave phases $\phi_k^0$. The second-order and forth-order invariants $I_1$ and $I_2$ do not depend on $\phi_k^0$ values while the firth-order invariant $I_3$ depends on their symmetric combination $\xi = \sum_{n=0}^{4} \phi_{2n}^0$ only. Therefore six-order expansion (3) is the simplest one. The fourth-order expansion does not depend on $\xi$ value and the fifth-order one has no global minimum. Note also that energy (3) does not depend on the other four mutually orthogonal linear combinations of phases $\phi_k^0$ [18].

Starting from this point we deviate from the conventional theory. To calculate the coordinates $\mathbf{r}_i$ of the cluster centers we represent the amplitudes in the form $\rho_k = \frac{1}{S}\sum_{n=1}^{N} \exp(-i\mathbf{b}_k\mathbf{r}_n)$, related to the cluster distribution, and substitute them into Eq. (3). Here N is the number of clusters and S is the area of the structure. Then the coordinates $\mathbf{r}_i$ can be found by minimizing Landau free energy (3). The solution of this problem in its general form is cumbersome and contains additional information about the phases others than decagonal quasicrystalline one, which are out of the frame of the present work. To illustrate the main conclusions of the model concerning decagonal QC formation we limit in the following consideration Landau expansion (3) to the effective free energy of the direct transition from the isotropic state to the decagonal quasicrystalline phase. For that goal we use the fact that ten-fold symmetry of this phase leads to the equality of all $|\rho_k|$ values. Using the parametrization (18) of density wave phases $\phi_k^0$ in terms of Goldstone variables **u** and **v** the function $\delta\rho(\mathbf{R})$ is then rewritten in the following real form:

$$\delta\rho(\mathbf{R}) = 2\rho_\Delta \sum_{n=0}^{4} \cos(\mathbf{b}_{2n}\mathbf{R} + \phi_{2n}^0), \quad (4)$$

where $\phi_{2n}^0 = \mathbf{b}_{2n}\mathbf{u} + \mathbf{b}_{2n}^\perp \mathbf{v} + \xi/5$, $\rho_\Delta = |\rho_k|$, $\mathbf{b}_n^\perp = b^0 \langle \cos(n3\pi/5), \sin(n3\pi/5) \rangle$. Accordingly, free energy (3) is simplified to the form:

$$F_q = A_1 \rho_\Delta^2 + A_2 \rho_\Delta^4 + A_3 \rho_\Delta^5 + A_4 \rho_\Delta^6, \quad (5)$$

where the coefficients $A_i$ are linearly dependent on the coefficients $F_i$ of free energy (3). In addition, since $I_3$ transforms to $2\rho_\Delta^5 \cos(\xi)$, the coefficient $A_3$ depends on $\xi$ value: $A_3 = 2F_4 \cos(\xi)$. The minimization of (5) with respect to the variable $\xi$ selects the structures with particular discrete values of $\xi$, which correspond to decagonal QC's. Namely, in the case $F_4>0$ the free energy minimum corresponds to $\cos(\xi) = -1$ and, consequently, to the QC phase with $\xi=\pi+2\pi\xi_0$, where $\xi_0$ is integer. In the case $F_4<0$, decagonal QC phase corresponds to the free energy minimum with the variable $\xi$ equal to $2\pi\xi_0$ and $\cos(\xi)=1$. Both cases can be considered separately in the frame of Eq. (5) with the coefficient $A_3$ taken to be equal to $\pm 2F_4$ in a function of $\cos(\xi)$ sign. In what follows we study in detail the case of $A_3$ negative, which is energetically favorable for $\xi=2\pi\xi_0$. The consideration of the opposite sign case is quite similar.

The explicit form of the effective order parameter (OP) $\rho_\Delta$ is deduced from the fact that the order parameter (OP) should be totally symmetric in the low-symmetry phase [23]. Therefore the OP of the isotropic-to-decagonal phase transition is expressed as the following average of $\rho_k$ amplitudes:

$$\rho_\Delta = \frac{1}{5S} \sum_{n=0}^{4} \sum_{i=1}^{N} \cos(\mathbf{b}_{2n}\mathbf{r}_i + \phi_{2n}^0). \quad (6)$$

Note that OP $\rho_\Delta$ given by Eq. (6) is a nonlinear bound-valued function. It is limited by its negative minimal and positive maximal values: $\rho_\Delta^{min} \leq \rho_\Delta \leq \rho_\Delta^{max}$. This constraint is caused by the dependence of the OP value on the cluster coordinates $\mathbf{r}_i$. Similar dependence of the OP value on atomic coordinates was introduced for the first time in [24] to explain the peculiarities of reconstructive phase transitions between complex crystalline phases. The maximal value limit $\rho_\Delta = \rho_\Delta^{max}$ of Eq. (6) is attained when all the maxima $(\delta\rho(\mathbf{r}_i)>\rho_{cf})$ of Eq. (4) coincide with the centers $\mathbf{r}_i$ of all the clusters, where the cut-off parameter $\rho_{cf}$ is determined by the surface concentration $C_s=N/S$ of nodes occupied by clusters. The global minimum of the free energy as a composite function of cluster centers coordinates $F_q = F_q(\rho_\Delta(\mathbf{r}_i))$ is then attained in the same point $\rho_\Delta = \rho_\Delta^{max}$, and $\mathbf{r}_i$ values corresponding to the free energy minimum are equal to those obtained by the maximization of Eq. (6). In the phase diagram of free energy (5) (Fig.1), the region of stability of the state with $\rho_\Delta = \rho_\Delta^{max}$ is located to the left from the dotted line $A_1 = -2A_2\rho_\Delta^2 - 5/2 A_3\rho_\Delta^3 - 3A_4\rho_\Delta^4$, with $\rho_\Delta = \rho_\Delta^{max}$ (see Fig. 1(a)). In this entire extended region the free energy minimization is equivalent to the maximization of effective OP (6). It is easy to see that this region on the phase diagram corresponds to the condition $\partial_{\rho_\Delta} F_q(\rho_\Delta^{max}) \leq 0$ as it is illustrated in Fig. 1(b).

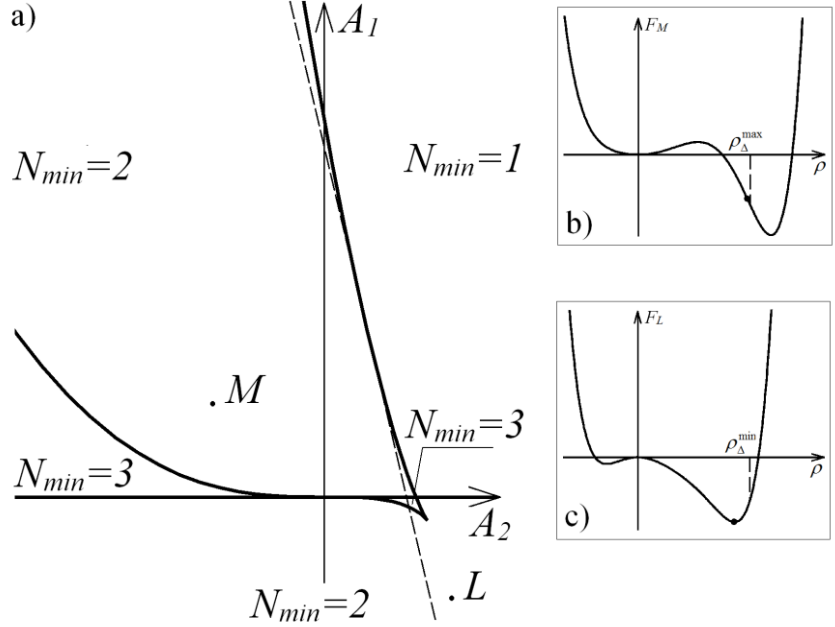

FIG. 1. Phase diagram corresponding to effective free energy (5). Panel (a) Regions of stability for different states of the system. Solid lines divide the diagram into regions with different numbers $N_{min}$ of the free energy $F_q(\rho_\Delta)$ minima. The minimum $\rho_\Delta = 0$ (isotropic liquid state) exists in the region $A_1 > 0$ only. The limit phase $\rho_\Delta = \rho_\Delta^{max}$ exists to the left from the dotted line. In this extended region the free energy minimization is equivalent to the maximization of the effective order parameter. The panels (b) and (c) show the form of Eq. (5) in regions M and L located to the left and to the right from the dotted line, respectively. Small solid circle shows the equilibrium value of the effective OP in the both cases.

For the state of lower symmetry, with a general value of $\rho_\Delta^{min} < \rho_\Delta^0 < \rho_\Delta^{max}$, stable in the region to the right of the dotted line in Fig. 1(a), the determination of $\mathbf{r}_i$ coordinates in the frame of effective potential (5) is not possible. Indeed, in this state, $F_q(\rho_\Delta)$ has a global minimum in the point $\rho_\Delta^0$, with $\rho_\Delta^{min} < \rho_\Delta^0 < \rho_\Delta^{max}$ (Fig. 1(c)), and the equality $\rho_\Delta(\mathbf{r}_i) = \rho_\Delta^0$ is not sufficient to determine the $\mathbf{r}_i$ coordinates.

The thermodynamics of the decagonal QC state formation and its relation to cluster centers positions presented above is the first step of the unified theory proposed. On the next step let us consider the structure which results from the maximization of the OP given by Eq. (6). As we discussed above, to maximize Eq. (6) the cluster centers should be located in the highest maxima of function (4). To determine the coordinates $\mathbf{R} = \mathbf{r}_j$ of these clusters we use the following system:

$$\mathbf{b}_{2n}\mathbf{r}_j + \Delta\phi_{2n}^j + \phi_{2n}^0 = 2\pi N_{2n}^j, \qquad (7)$$

where $N_{2n}^j$ are integers; n=0,1,2,3,4; and $\Delta\phi_{2n}^j$ are small phase deviations. The deviations appear because of the fact that individual waves in superposition (4) are mutually incommensurate. Thus, even in the highest maxima of function (4) the phases of individual waves are slightly different from the $2\pi N_{2n}^j$ value. Using the fact that $\sum_{n=0}^{4}\Delta\phi_{2n}^j = 0$ we parameterize the small phase deviations $\Delta\phi_{2n}^j$ by two 2D vectors $\Delta\mathbf{r}_j$ and $\mathbf{r}_j^\perp$: $\Delta\phi_{2n}^j = \mathbf{b}_{2n}\Delta\mathbf{r}_j + \mathbf{b}_{2n}^\perp \mathbf{r}_j^\perp$. After that the solution of Eqs. (7) takes the following form:

$$\mathbf{r}_j + \Delta\mathbf{r}_j = \sum_{i=0}^{4} N_{2i}^j \mathbf{a}_i - \mathbf{u}, \tag{8}$$

$$\mathbf{r}_j^\perp = \sum_{i=0}^{4} N_{2i}^j \mathbf{a}_i^\perp - \mathbf{v}, \tag{9}$$

$$\xi_0 = \sum_{i=0}^{4} N_{2i}^j. \tag{10}$$

where $\mathbf{u}$ and $\mathbf{v}$ are homogeneous phonon and phason shifts of the structure, respectively; $\mathbf{a}_i = \frac{4\pi}{5b_0}(\cos(i2\pi/5),\sin(i2\pi/5))$, and $\mathbf{a}_i^\perp = \frac{4\pi}{5b_0}(\cos(i6\pi/5),\sin(i6\pi/5))$, $i=0,1,2,3,4$. In the vicinity of any high maximum $\mathbf{r}_j$ function (4) can be approximated by the expression $\delta\rho(\mathbf{R}) \approx 2\rho_\Delta\left(5 - \sum_{n=0}^{4}(\Delta\phi_{2n}^j)^2/2\right)$. In this approximation the maximization of the $\delta\rho(\mathbf{R})$ function with respect to the vector $\mathbf{R}$ components yields

$$\mathbf{R} = \mathbf{r}_j = \sum_{i=0}^{4} N_{2i}^j \mathbf{a}_i - \mathbf{u} \tag{11}$$

and

$$\delta\rho(\mathbf{r}_j) \approx 10\rho_\Delta\left(1 - (\mathbf{r}_j^\perp)^2 b_0^2/4\right). \tag{12}$$

Therefore the deviations $\Delta\mathbf{r}_j$ in Eq. (8) are small and $|\Delta\mathbf{r}_j| \ll |\mathbf{r}_j^\perp|$ for any high maximum of Eq. (4).

Let us stress that Eqs. (9-11) coincide completely with the conventional equations which express in high-dimensional crystallography the projection of a planar decagonal quasilattice from the 5D space. The values of density function (4) in the region $\delta\rho(\mathbf{R}) > 3.08\rho_\Delta$ are presented in Fig. 2 and superimposed with the projections of the 5D space lattice nodes (small circles). Fig. 2 shows an excellent coincidence of the highest maxima of $\delta\rho(\mathbf{r}_i)$ with the projections of the 5D nodes. All maxima of the density function in the region $\delta\rho(\mathbf{r}_j) > 3.08\rho_\Delta$ are uniquely indexed by integer indices $N_{2n}^j$. At the scale chosen in Fig. 2 the deviations $\Delta\mathbf{r}_j$ between the maxima positions and the 5D nodes projections are extremely small.

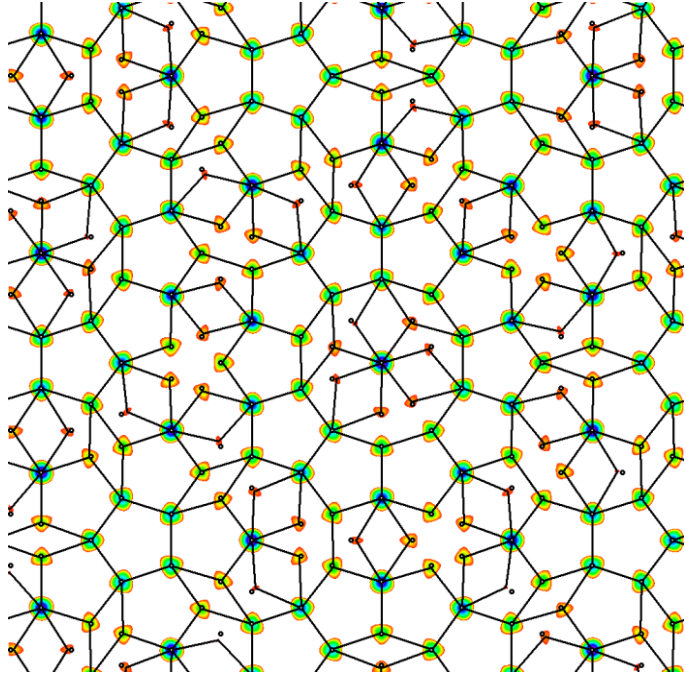

FIG. 2. (color online) Values of density function (4) (with $\xi_0=1$) in the region $\delta\rho(\mathbf{R}) > 3.08\ \rho_\Delta$ superimposed with the projections of the corresponding 5D space lattice nodes. Color variation from red to violet corresponds to the $\delta\rho(\mathbf{R})$ value variation from $3.08\rho_\Delta$ to $10\rho_\Delta$. Vertices of the quasicrystalline tiling coincide with the $\delta\rho(\mathbf{R})$ function maxima. Small circles representing the projections of the 5D space lattice nodes are calculated using Eq. (11).

The condition $\delta\rho(\mathbf{R}) > 3.08\rho_\Delta$ is equivalent to the certain choice of surface concentration of nodes $C_s$ for which the tiling obtained is very close to the experimental distribution of cluster centers in classical metallic QC $Al_{65}Cu_{20}Co_{15}$ [11,13]. However, the tiling in Fig. 2 shows small deviations from the experimental structure. Minimization of the free energy which takes into account the constraints imposed by the internal structure of clusters makes the tiling identical to the experimental cluster distribution. Corresponding modification of the classical crystallization theory is presented in the next section.

### III. MINIMIZATION OF THE LANDAU FREE ENERGY WITH THE CONSTRAINTS IMPOSED BY THE INTERNAL STRUCTURE OF THE CLUSTERS

Let us now analyze in detail the QC tiling in Fig. 2 in the frame of cluster approach. We will stay in the frame of initial formulation of the approach proposed by Burkov [11] and Yamamoto [13] for classical metallic QC $Al_{65}Cu_{20}Co_{15}$. We will compare the main characteristics of the tiling with the typical features of the $Al_{65}Cu_{20}Co_{15}$ clusters and their relative positions. The tiling presented in Fig. 2 consists of stars, truncated stars, thin and thick rhombuses. Up to small deviations $\Delta \mathbf{r}_j$ the distance between nearest neighbors in the tiling is $\tau = (\sqrt{5}+1)/2$ times smaller than the tiling edge length. Besides, the short diagonal in a thick rhombus is approximately 1.17 times longer than the tiling edge. On the other hand, the distances between $Al_{65}Cu_{20}Co_{15}$ clusters (proposed in [25] and used in both [11] and [13]) can be equal to either 2nm or 1.2nm. The ratio of these distances is also equal to $\tau$. For all other distances between the cluster centers the motives defined by the clusters are simply not coherent. Thus, the clusters considered can not be located in *all* nodes of the tiling presented in Fig. 2. The clusters situated in the nodes related by the short diagonal of thick rhombuses can not overlap in a regular way. Consequently, in each couple of the corresponding clusters, one of two clusters should be excluded. Similar feature characterizes Burkov's cluster model, in which only a sub-ensemble of rhombic Penrose tiling nodes is decorated with clusters. The explicit form of this sub-ensemble was found in [13] (see also Fig. 2(b) in this work). More detailed TEM image analysis performed in the same work has also shown that to understand the $Al_{65}Cu_{20}Co_{15}$ structure in the frame of cluster approach it is necessary to decorate with clusters PPQ (see Fig. 2(a) in [13]) and not rhombic Penrose tiling. The tiling presented in Fig. 2 is very close to PPQ. The only difference between two tilings is the absence of thick rhombuses in PPQ because of the absence of one of two nodes related by the distance approximately 1.17 times longer than the tiling edge (as it is discussed above).

The analysis of tiling decoration by Burkov and Yamamoto, together with the analysis of geometrical difference between PPQ and the tiling in Fig. 2 makes it clear that in the frame of cluster approach the minimization of Landau free energy (5) (and, consequently, the maximization of amplitude (6)) must be constrained. Conditional minimum of the free energy must take into account geometrical limitation on the quasilattice nodes occupation imposed by the internal organization of clusters. This limitation forbids simultaneous location of two clusters in two positions related by one of ten vectors $\mathbf{S}_i$ corresponding to short diagonals of thick rhombuses. Then, the way of cluster location in maxima of function (4) which minimizes free energy (5) is the following: *in the coherent cluster distribution the center of a given cluster corresponds to the amplitude maximum which is higher than all neighboring maxima related to the given one by vectors $\mathbf{S}_i$*. Fig. 3 presents the ideal PPQ obtained by constrained minimization of free energy (5). Nodes of PPQ correspond to the maxima of function (4). Vectors $\mathbf{S}_i$ are oriented from the highest maxima of (4) occupied by the clusters to lower maxima excluded by the constrained minimization of the free energy.

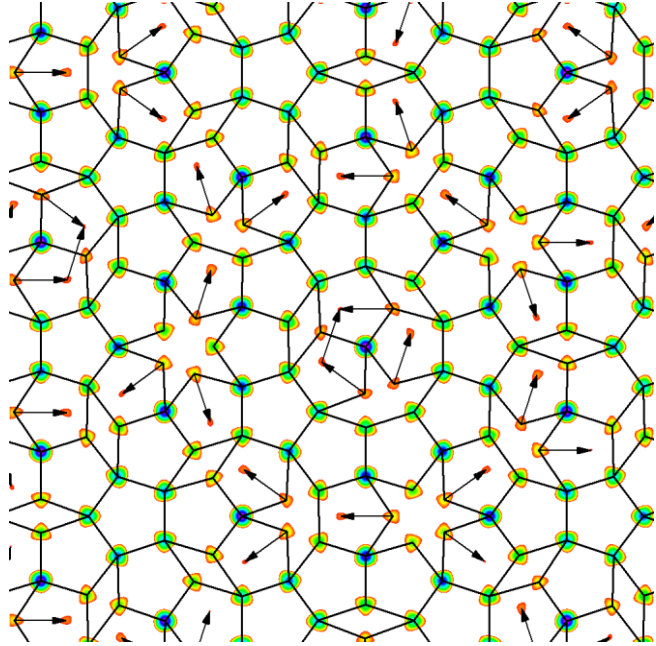

FIG. 3. (color online) Construction of the pentagonal Penrose quasilattice (PPQ) in the frame of the Landau crystallization theory constrained by the cluster organization limitation. Regions with the values $\delta\rho(\mathbf{R})>3.08\ \rho_\Delta$ are presented. Color code is the same as in Fig. 2. Each QL node corresponds to the amplitude maximum which is higher than all neighboring maxima related to the given one by vectors $\mathbf{S}_i$. Explicit form of $\mathbf{S}_i$ vectors and the details of constrained free energy minimization are discussed in the text.

Let us illustrate the relation between the PPQ construction algorithm proposed and the traditional methods of the N-d crystallography. We have already mentioned that maxima of function (4) associated in the proposed approach with the cluster centers are indexed with integer indices $N_{2n}^{j}$. Due to this fact, the vectors $\mathbf{S}_i$ are parallel projections of the 5D translations (up to small deviations $\Delta \mathbf{r}_j$). All of them are equivalent by symmetry to the translation $\mathbf{Z}_1$ =<-1,-1,0,2,0>, (full set of these vectors is obtained from $\mathbf{Z}_1$ by cyclic permutation and inversion of its components). Then, taking into account Eq. (10) and Eq. (12) the constrained energy minimization algorithm for the PPQ construction can be reformulated in the following way: *1) The lattice is constructed from nodes with the same fixed integer value of $\xi_0$. 2) The 5D node is projected onto the QL if the neighbors of the node related to it by the vectors $\mathbf{Z}_i$ and characterized by smaller $|\mathbf{r}_j^{\perp}|$ values are absent.* This 5D crystallography formulation of the proposed PPQ construction algorithm is perfectly equivalent to the conventional projection method [26] for the PPQ. It is easy to see that point 2) of the proposed algorithm for 5D nodes selection is equivalent to the conventional condition that all perpendicular coordinates $\mathbf{r}_j^{\perp}$ of a node find themselves in the projection window, which has the form of a regular decagon with the distance between opposite sides equal to the length of the vector $\mathbf{Z}_1$ perpendicular projection. It is also interesting to note that the PPQ constructed using constrained maximization of function (4) presented in Fig. 3 is the first QL in a series of self-similar QL's with successively longer tiling edges. All of them are also constructed using the constrained minimization of the same free energy (or maximization of the same $\delta\rho(\mathbf{R})$ function) algorithm. If we change the constraint imposed to energy (5) minimization by choosing the lengths of $\mathbf{S}_i$ vectors $\tau$ times longer (they become in this case parallel projections of the 5D translations equivalent by symmetry to the translation <1,-1,-1,1,0>), the proposed algorithm results in the PPQ with the edges $\tau$ times longer. For successive QL's in this series the ratio of the maximal deviation $\Delta\mathbf{r}_j$ and the QL edge length decreases progressively, and the positions of maxima of amplitude function (4) become closer and closer to the positions of projections of the 5D space nodes. Note also that for any QL the node density is $\tau^2$ times smaller than for the preceding QL in the series. The $C_s$ values for all QL's are completely determined by the constrained energy minimization procedure (and are not chosen arbitrarily as it is the case of the tiling in Fig. 2 obtained by unconstrained minimization).

## IV. CONCLUSION

In conclusion, we would like to stress that the proposed unified theory combining the Landau theory of crystallization and the cluster approach to quasicrystals permits to describe in the same frame assembly thermodynamics and structure formation for many different QL's. The algorithm proposed and illustrated above on the example of the decagonal QC, gives particularly simple description (by means of constrained minimization of the free energy dependent on a single irreducible OP) and easy node coordinates determination in the following cases:  i) the octagonal QL formed by squares and rhombuses [27]; ii) the Amman-Mackay 3D tiling [28] composed of oblate and prolate rhombohedra; iii) the dodecagonal tiling formed by squares, regular triangles and oblate hexagons [29]. Similar to the PPQ case discussed in detail in the present work, the ideal QL's in these cases are obtained by the free energy minimization constrained by a limitation which forbids simultaneous location of two identical clusters in two positions related by a specific symmetric star of vectors. The constrained free energy minimization approach constitutes a novel physical justification for the purely crystallographic concept of QL construction which is based on the projection of high-dimensional space nodes using projection window. The structures obtained by the projection method can also be obtained in the frame of the Landau theory of crystallization constrained by simplest limitations on the relative positions of structural units in QC phases due to internal structure of clusters.